# Observation of short-period helical spin order and magnetic transition in a non-chiral centrosymmetric helimagnet

*Bei Ding, Jun Liu, Hang Li, Jinjing Liang, Jie Chen, Zefang Li, Xue Li, Xuekui Xi, Zhenxiang Cheng, Jianli Wang, Yuan Yao[*] & Wenhong Wang[*]*

Dr. B. Ding, Dr. J. Liu, Dr H. Li, J. J. Liang, Z. F. Li, X. Li, Dr. X. K. Xi, Dr Y. Yao, Prof. W. W. Wang
Beijing National Laboratory for Condensed Matter Physics,
Institute of Physics, Chinese Academy of Sciences,
Beijing 100190, China

E-mail: wenhong.wang@iphy.ac.cn; yaoyuan@iphy.ac.cn

Dr. J. Chen, Prof. W. W. Wang
Songshan Lake Materials Laboratory
Guangdong 523808, China

J. J. Liang, Z. F. Li, X. Li, Prof. W. W. Wang
University of Chinese Academy of Sciences
Beijing 100049, China

Prof. Z. X. Cheng, Prof. J. L. Wang
Institute for Superconducting and Electronic Materials, Innovation Campus, University of Wollongong, Squires Way, North Wollongong, NSW 2500, Australia

**Abstract:**

The search for materials exhibiting nanoscale spiral order continues to be fuelled by the promise of emergent inductors. Although such spin textures have been reported in many materials, most of them exhibit long periods or are limited to operate far below room temperature. Here, we present the real-space observation of an ordered helical spin order with a period of 3.2 nm in a non-chiral centrosymmetric helimagnet MnCoSi at room temperature via multi-angle and multi-azimuth approach of Lorentz transmission electron microscopy (TEM). A magnetic transition from the ordered helical spin order to a cycloidal spin order below 228 K is clearly revealed by in situ neutron powder diffraction and Lorentz TEM, which is closely correlated with temperature-induced variation in magneto-crystalline anisotropy. These results reveal the origin of spiral ordered spin textures in non-chiral centrosymmetric helimagnet, which can serve as a new strategy for searching materials with nanoscale spin order with potential applications in emergent electromagnetism.

# 1. Introduction

The ordering of magnetic moments in non-collinear or non-coplanar textures is known as a source of plentiful and intriguing phenomena. Among these, emergent inductors that utilize emergent electric fields generated by the current-induced motion of spiral spin texture have recently received growing attention for their potential applications in realizing dramatic miniaturization of inductance elements[1, 2]. Regardless of the direction of the helical plane, an emergent electric field can be induced in non-collinear spin texture, including incommensurate helixes and cycloids. Such unique inductance has been recently reported in bulk centrosymmetric magnetic materials $Gd_3Ru_4Al_{12}$[3] and $YMn_6Sn_6$[4] with a short period ($\leq$ 3 nm) of helical spin state. Whether these intriguing electromagnetic phenomena are generic and common to other spiral spin textures, the quest for materials hosting short-period spin textures such as helixes and cycloids continues to be fuelled by the promise of emergent inductors.

There are several origins of the helical spin structure proposed in helimagnets. One of them is in chiral noncentrosymmetric crystals, such as MnSi[5, 6], FeGe[7], $Fe_{1-x}Co_xSi$[8, 9], $Cr_{1/3}NbS_2$[10, 11] and $CsCuCl_3$[12], in which inversion symmetry breaking induces long-period helical spin order stabilized by Dzyaloshinskii-Moriya (DM) interactions[13, 14]. Another origin of the helical spin structure is in the non-chiral centrosymmetric crystals called Yoshimori-type helimagnets[15], where the competition of exchange interactions between the nearest neighbour and next nearest neighbour results in the helical spin order[16], which has been reported in $MnO_2$[15], rare-earth metals Dy[17], Tb[18] and Ho[19, 20]. One possible path to the discovery of new centrosymmetric magnets that can host nanoscale spiral spin order is to revisit noncubic, noncollinear magnets that have been reported to be ferromagnetic above room temperature.

Recently, rich magnetic features with unique noncollinear magnetism beyond room temperature were experimentally observed in XMnZ alloys (TiNiSi-type) [21, 22]. Among the isostructural TiNiSi-type compounds, MnCoSi has attracted particular interest due to large magnetoelastic effects accompanied by magnetic structure transitions[23]. Combined with magnetic measurements and neutron powder diffraction (NPD) data, nanoscale noncollinear incommensurate helical antiferromagnetic activity in MnCoSi at temperatures below 380 K has been widely reported[22-27]. More intriguingly, recent NPD studies in MnCoSi found that a nanometric cycloidal spin order may exist at lower temperatures[28, 29]. These exotic features in MnCoSi offer a good platform to study emergent inductors; however, real-space observations of nanoscale magnetic structures and related magnetic structure transitions in MnCoSi are still lacking.

Lorentz transmission electron miscopy (LTEM) is a powerful tool to visualize magnetic configurations in real space by imaging the phase shift induced by magnetization perpendicular to the incident electron beam. Although this technique has successfully revealed different kinds of helical-spin orders in various materials, such as the metal silicide family[30-32], rare-earth metals[33], FeGe[34], and intercalated 2H-type $NbS_2$[35-37], few characterizations of the short-period spiral in XMnZ (TiNiSi-type) alloys have been reported due to the weak phase signal of magnetic structures and restricted spatial resolution. Moreover, LTEM images are two-dimensional (2D) projections of the in-plane spin component integrated along the electron beams in an electron-transparent thin specimen lacking three-dimensional (3D) information, which is inadequate to identify the spin configuration of a magnetic texture. Thus, additional operations are required to reconstruct the 3D topography of spiral structures in magnets via LTEM.

In this communication, the multi-azimuth and multi-angle imaging approach of LTEM was applied to study the non-chiral centrosymmetric helimagnet MnCoSi (TiNiSi-type) to elucidate the internal 3D magnetic spiral configurations in real space. Combined with NPD data, the results unambiguously demonstrate that magnetic moments rotating in the *ab* plane of the crystal form pure helical orders propagating along the *c* axis with a period of approximately 3.2 nm at room temperature. With decreasing temperature, the helical orders gradually evolve into cycloids as the temperature decreases to 228 K, which means that the magnetic moments rotate in the *bc* plane, but the propagation vector persists in the *c* direction. The real-space observation of the exotic nanometric spiral in the metallic metamagnet MnCoSi not only opens an opportunity to distinguish the complex spin textures in magnets but also reveals the functionality of this material as a new paradigm for spintronics storage and emergent electromagnetic induction.

## 2. Results and Discussion

High-quality single crystalline MnCoSi was synthesized via the flux method for LTEM characterization (see the Experimental section). The crystal structure verified by X-ray diffraction (XRD) and NPD was an orthorhombic structure in space group *Pnma*, which agrees well with previous reports[38]. MnCoSi possesses four Mn, Co, and Si atoms in a unit cell, and all atoms occupy the *4c* (*x*, 1/4, *z*) crystallographic positions, as shown in **Figure 1**a. Previous NPD studies indicated the coexistence of four helixes at room temperature, i.e., Mn1-Mn3, Mn2-Mn4, Co1-Co3, and Co2-Co4, with a constant phase difference due to indirect double exchange coupling[24]. For clarity, only the moments of Mn atoms are highlighted in Figure 1a because they dominate the magnetic structure in MnCoSi ($\mu_{Mn}$= 2.2 $\mu_B$, $\mu_{Co}$= 0.3 $\mu_B$). The spins rotate in the *ab* plane, with the helical vector $k_z$ along the *c* axis.

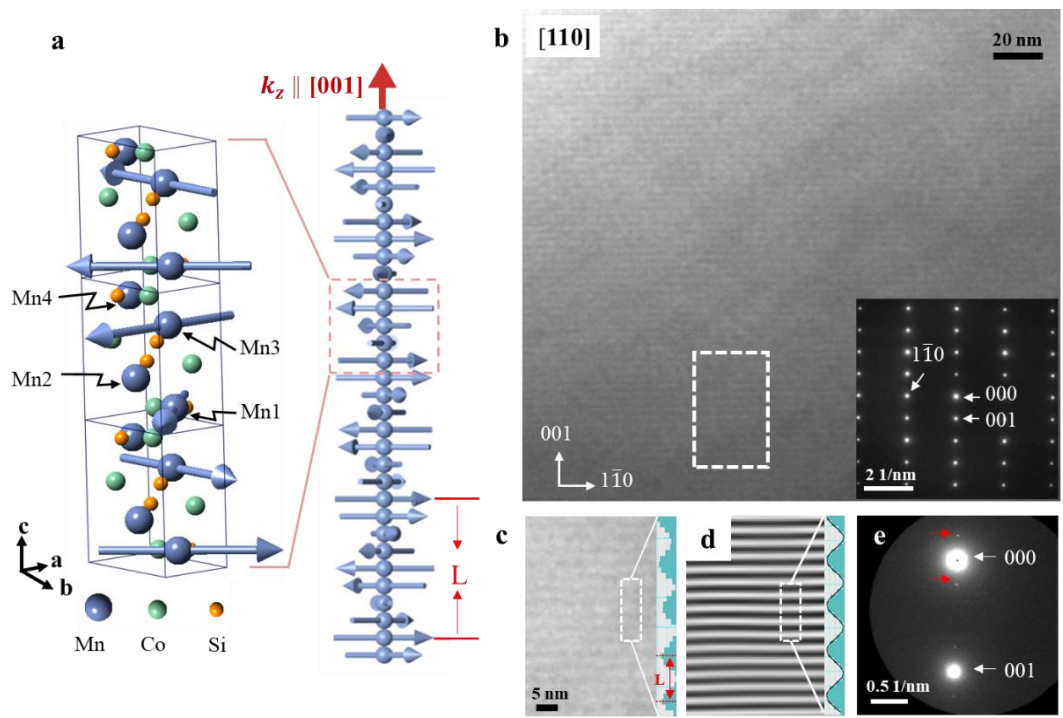

**Figure 1.** Crystal and magnetic structure of MnCoSi. a) Schematic representation of the crystalline structure and magnetic helical order of MnCoSi. The helical vector $k_z$ is along the [001] (*c* axis) in an orthogonal structure, while spins rotate in the *ab* plane. The screw pitch is *L*. b) Under-focused LTEM image and selected area electron diffraction (SAED) of magnetic configurations in the [110] MnCoSi specimen at 297 K without an external field. The defocus value is 2.8 μm. SAED was conducted at a camera length (CL) of 60 cm. c) and d) Magnified and Fourier filtered images of the area indicated in (b), respectively, where the magnetic contrasts appear as bright and dark stripes with a period of 3.4 nm. e) Large-magnified diffraction patterns show the satellite spots near the main reflections, which correspond to the magnetic spiral structure. The location of the satellites is also equal to the 3.2 nm periodic stripes. SAED was conducted at a CL of 780 cm.

Thin LTEM specimens along the [100], [110] and [010] zone axes were fabricated via focused ion beam (FIB) with the assistance of electron backscatter diffraction (EBSD) (see Figures S1 and S2, Supporting Information). Figure 1b displays an underfocused LTEM image of a [110] specimen under zero external field at 297 K, together with the selected area electron diffraction (SAED) patterns. The defocus value was 2.8 μm to ensure a sufficient spatial resolution. The repeating stripes normal to the [001] direction (*c* axis) clearly spread all over the observed region. A square area was selected and filtered by the Fourier method to highlight the magnetic contrast, as demonstrated in Figure 1c and 1d. The bright and dark stripes in the defocused images indicate the opposite directions of the in-plane magnetic moment[39]. The sinusoidal modulation of the line profile outlines the stripes with a period (*L*) of 3.4 nm along the [001] direction. The overfocused image also confirms the same period as the reversal magnetic contrast (Figure S3, Supporting Information). The period can also be distinguished as two satellite spots adjacent to the main reflections in the enlarged SAED patterns, as shown in Figure 1e. Moreover, similar stripes perpendicular to the *c* axis and with periods of approximately 3.6 nm also appeared in other lamellae with [100] and [010] zone axes (Figure S4, Supporting Information), confirming that spin textures are intrinsic to the material, even in bulk MnCoSi crystalline form.

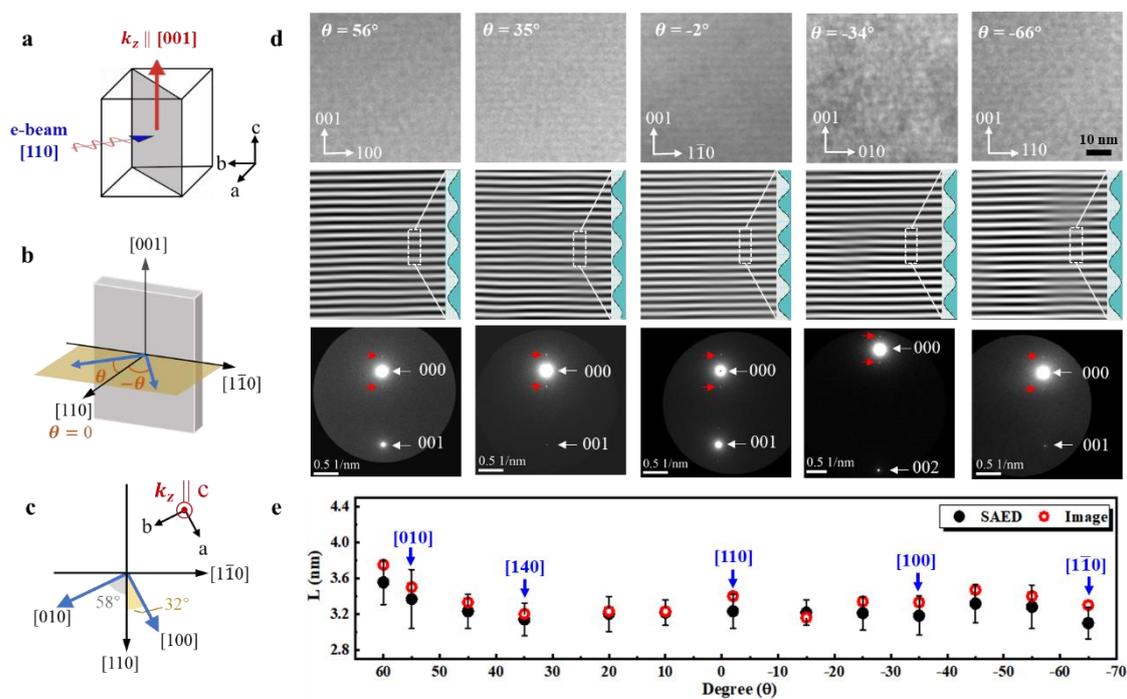

**Figure 2.** Angle dependence of magnetic spin textures. a) Schematic figure of the [110] specimen obtained

by LTEM observation, where the electron beam is injected along the [110] direction and a-b-c indicate the crystal lattice vectors. b) Experimental setup for serial-tilting imaging in which the specimen was rotated around the screw axis (*c* axis). c) Relative angles among different main zone axes in the *ab* plane. d) Under-focused images, Fourier-filtered images and corresponding SAED at various tilt angles, with satellite spots and contrast profile indications. Note that SAED was conducted at a CL of 780 cm except $\theta$ = -34° (CL = 470 cm). e) Measured spiral periods as a function of tilt angle, where some special crystal directions are denoted. The black scatters and red open circles represent the data from SAED and images, respectively. The error bar represents the standard deviation of SAED measurements.

To identify the magnetic configuration of stripes in detail, serially tilted images were acquired with a high-tilt holder. For this purpose, the *c* direction of a [110] crystalline specimen was deliberately placed along the rotation axis of the high-tilt holder to ensure that the image contrast always arose from the projected magnetic features in the *ab* plane during sample tilting. The initial image was the plane perpendicular to the [110] direction, and then the sample was tilted between ±70° to observe the change of magnetic contrast (**Figure 2**a-c). Figure 2d shows the defocused LTEM images, Fourier-filtered images and corresponding SAED patterns acquired at $\theta$ = 56°, 35°, 0°, -34°, and -66°, demonstrating the stable appearance of the periodic stripes in different sample orientations. Figure 2e unambiguously verifies the almost constant screw pitches upon tilting the sample. These equidistant stripes in different tilted images are consistent with the helical magnetic moments spinning in the *ab* plane, which invariably contribute to the bright and dark contrast because there are always magnetic components perpendicular to the electron beam, regardless of the rotation of the specimen around the helical axis (*c* axis). The serial-tilting characterization for the [100] and [010] samples showed the same phenomena (Figures S5 and S6, Supporting Information). These results strongly confirm that a robust fine helical magnetic order exists in the non-chiral metallic meta-magnet MnCoSi at room temperature, with a period comparable to that of $Gd_3Ru_4Al_{12}$[3], $YMn_6Sn_6$[4] and the rare-earth metal Dy[33] but smaller than that of the non-centrosymmetric helimagnets, such as $Cr_{1/3}NbS_2$[35] and $Mn_{1/3}NbS_2$[37].

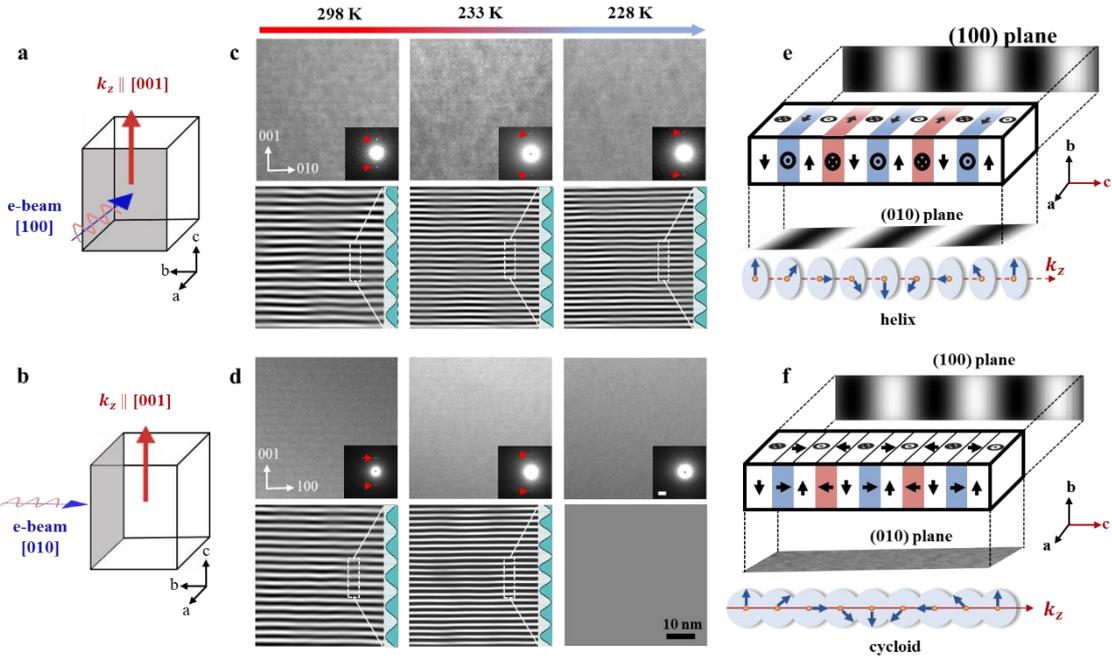

**Figure 3.** Temperature dependence of magnetic spin texture. a) and b) Schematic figures of thin plates in real space indexed with the [100] and [010] zone axes. The sample was fabricated from a single crystal with a propagation vector $k_z$ along the [001] direction. c) and d) Representative LTEM images, Fourier-filtered images and SAED patterns of [100]- and [010]-oriented specimens at different temperatures, respectively, where the contrast profile and satellite spots from the magnetic orders are denoted. The scale bar in the SAED pattern is 0.1 1/nm. e) and f) Schematics of helical and cycloidal spin order propagating along the *c* axis and the corresponding simulated LTEM images projected on different imaging planes. For the helical order, the magnetization rotates in the *ab* plane, while for the cycloidal structure, the magnetic moments lie on the *bc* plane.

Having visualised the helical spin order of MnCoSi at room temperature, we now investigate its spin texture at low temperatures. [010]-, [110]- and [100]-oriented lamellae were cooled to 133 K in an in situ cooling holder to analyse the phase change of the magnetic structures in MnCoSi. **Figure 3** displays the changes in the LTEM images and SAED patterns of the [100] and [010] samples. At room temperature, stripes appeared in the images of both the [100] and [010] samples, accompanied by satellite spots in the SAED patterns, as described above. When the temperature decreased below 228 K, the stripes in the [100] sample remained with a decreased period, as confirmed by the increase in the distance between satellite diffractions. However, the magnetic contrast of the stripes disappeared from the [010] lamella synchronously with the loss of satellite spots (Figure 3d and Figures S7 and S8, Supporting Information). The continuous dimming of the satellite diffractions along the [010] direction is clearly recorded in Videos S1 and S2. Notably, a new magnetic feature distinct from the initial helical order emerges at approximately 228 K. Based on previous investigations[40], a possible magnetic order named the cycloidal structure of MnCoSi below 228 K is shown in Figure 3e and 3f with the different simulated LTEM images in the two vertical [100] and [010] view directions. Unlike the helical spin order in which bright/dark stripes are always observed in the images during specimen rotation around helical vector $k_z$, cycloids with the spins rotating in the *bc* plane containing screw vector $k_z$ represent stripe contrast only in the *bc* plane and no magnetic features in the *ac* plane (Figure 3f). This magnetic structure is in exact agreement with the LTEM observations of MnCoSi at lower temperature. Therefore, the different contrast between the [100] and [010] samples at 228 K readily reveals that the helical structure is converted to a stable cycloidal order when the temperature is below the critical point.

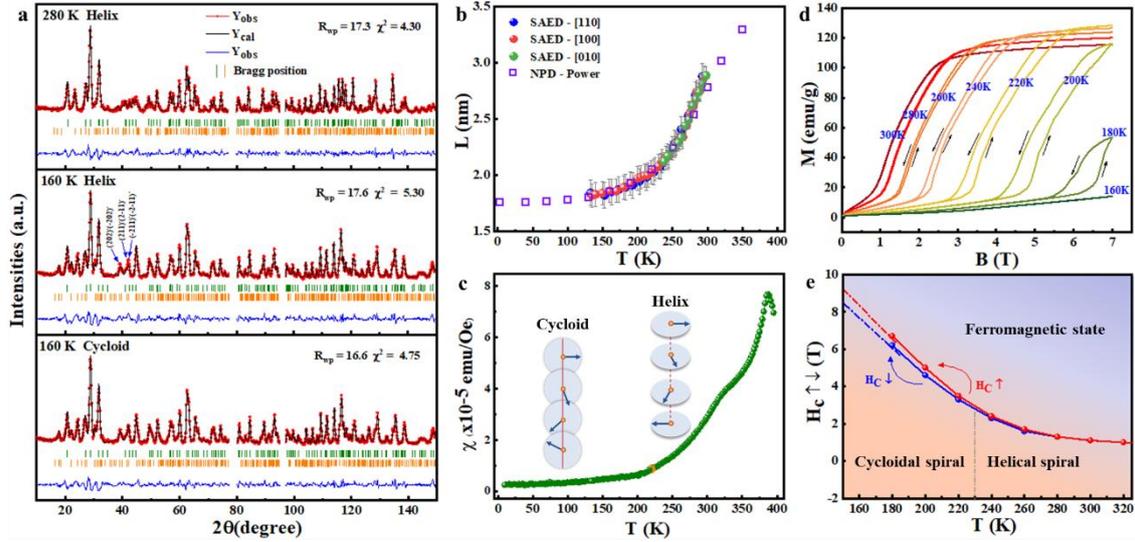

**Figure 4.** NPD patterns and the related magnetic transition. a) Temperature-dependent NPD patterns. Experimental (red) and calculated NPD patterns (black) and their difference profiles (blue) at selected temperatures are shown. Vertical lines indicate the peak positions for the nuclear (green) and magnetic (orange) reflection of the MnCoSi phase. b) Plots of the period $L$ of the [100], [110] and [010] specimens at various temperatures, measured from the SAED patterns and NPD data. The error bars were deduced from the tilted SAED data in Fig. 2e. c) Temperature dependence of ac magnetic susceptibility of MnCoSi polycrystalline (H = 10 Oe, f = 1000 Hz). The inset represents the spin configuration at different temperatures. The magnetic transition temperature is marked by the orange dot. According to LTEM observations, the cycloidal and helical magnetic states are separated. d) Isothermal magnetization of bulk polycrystalline MnCoSi between 160 K and 300 K at 20 K intervals. The arrows indicate the direction of field ramping. e) Proposed magnetic phase diagram of bulk MnCoSi constructed from M-H loops, where the lower line corresponds to $H_C\downarrow$ and the upper line $H_C\uparrow$. According to LTEM observations, the cycloidal and helical magnetic states are separated.

Further insight into the magnetic structure was accomplished through NPD measurements. Data were taken from 450 K to 5 K in a zero magnetic field. Representational refinement of the NPD data at 280 K and 160 K is shown in **Figure 4a**. With the temperature decreasing below 385 K, some magnetic peaks start apparent in the low-angle range of the diffractogram. For T= 280 K, the NPD patterns can be fitted well by the helical magnetic structure model, for which the spin arrangement of Mn and Co atoms lying in the *ab* plane achieves an incommensurate propagation vector $\boldsymbol{k} = (0, 0, 0.2721(8))$. Some typical magnetic Bragg reflections of $(202)^-$, (-

202)⁻, (211)⁻, (2-11)⁻, (-211)⁻ and (-2-11)⁻ are indexed. However, when the temperature decreases to 160 K, the simulated patterns of the helical magnetic model and observed patterns represent a small discrepancy, which further indicates that a new magnetic structure emerges at low temperature. Assisted by symmetry arguments[41], a cycloidal magnetic model is performed to fit the data, which represents lower values of R$_{wp}$ and $\chi^2$ and a better fitting of the main magnetic diffraction peaks. Specifically, the magnetic moments of this cycloidal spiral structure lie in the *bc* plane and roll along the *c* axis with an incommensurate ***k*** = (0, 0, 0.3678(6)), which corresponds well with the observation of the LTEM images at low temperature. Figure 4b shows how the spiral spin order period *L* varies with temperature T. The period *L* of the [010], [100] and [110] lamellae at different temperatures was measured from the scattering spots in the SAED patterns. For the [100] and [110] samples, upon decreasing the temperature, *L* represents a sharp decrease above 228 K and gradually remains constant up to 1.83 nm near 133 K, which corresponds well with the variation in NPD data. Interestingly, a break point of *L* in the [010] sample (green dots) occurs, which indicates that the magnetic structure changes at approximately 230 K. The period variation of spiral spin textures in the bulk MnCoSi polycrystals was further studied and confirmed by magnetic measurements, as shown in Figure 4c. The magnetization decreases sharply from 385 K to 228 K and gradually remains constant at 5 K. Figure 4d shows the isothermal magnetization vs applied field for polycrystalline MnCoSi. The antiferromagnetic spiral to ferromagnetic state transition was clearly observed from 300 K to 180 K. A stronger magnetic field should be applied at a lower temperature (<160 K) to induce a magnetic transition, yielding results similar to those of previous reports[42, 43]. By combining the LTEM images and magnetization measurements, we then plotted the magnetic phase boundaries of magnetic configuration on the basis of the metamagnetic critical field *H$_c$*, which is defined as the field at which the magnetic moment reaches 50% of its saturation value, as presented in Figure 4e. The critical field for increasing ($H_c \uparrow$) and decreasing ($H_c \downarrow$) fields reveals a small difference, indicating the hysteresis in this symmetric helimagnet. Clearly, the helical order in MnCoSi should be stable over an extremely wide temperature vs magnetic field range, and a new cycloid phase emerges at low temperatures.

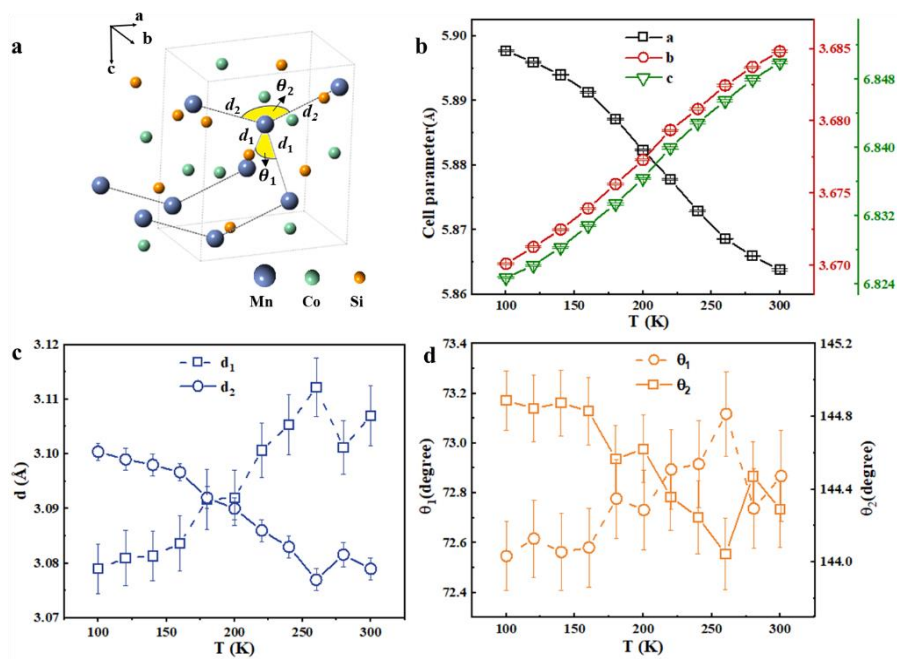

**Figure 5** Temperature dependent crystal parameter measured on MnCoSi powders using XRD. a) Definition of the two shortest Mn-Mn distance $d_1$ and $d_2$ and the Mn-Mn-Mn bond angles $\theta_1$ and $\theta_2$, respectively. The blue, green and orange ball depict the Mn, Co and Si atoms, respectively. b) Cell parameters *a*, *b*, *c*, c) Mn-Mn distance and d) Mn-Mn-Mn bond angles as a function of temperature via the refinement of XRD patterns. As temperature varied, the tiny change occurs in cell parameters, Mn-Mn distance d and Mn-Mn-Mn bond angles $\theta$.

Our experimental results showed that a helical spin order of 3.2 nm indeed exists in the centrosymmetric helimagnet MnCoSi at room temperature and transformed into cycloidal spin order below 228 K. In conjunction with the magnetization, we discuss a possible physical mechanism of magnetic transition. Based on the first-principles calculation and experimental statistics, the magnetic ground state of Mn-based orthorhombic alloy with space group *Pnma* is closely related to the Mn-Mn bond distance and bond angles [44]. For orthorhombic MnCoSi alloy, its Mn-Mn separation locates on the border of FM and AFM, which can easily be perturbed by crystal fields and magnetoelastic effects in a manner dependent on the temperature and magnetic field. **Figure 5** shows the temperature evolution of cell parameters (*a, b, c*), two nearest-neighbour Mn-Mn distances ($d_1$, $d_2$) and Mn-Mn-Mn bond angles ($\theta_1$, $\theta_2$), obtained from Rietveld refinement of XRD data measured from 300 K to 100 K (Figure S9 and Table S1, Supporting Information). With decreasing temperature, the lattice parameters *b* and *c* typically decrease, while the lattice parameter *a* shows a negative thermal expansion effect, as reported in previous works [23, 42]. The interatomic distance $d_1$ increases with temperature while the bond angle enclosed by $d_1$, $\theta_1$, decreases. For $d_2$ and $\theta_2$, it is the other way around. The variation of Mn-Mn distance and bond angles makes variation of the exchange interaction between magnetic Mn atoms which further results in the periodicity of spiral magnetic structure decreases. On the other hand, we notice that a crossover of plots $d_1$, $d_2$ appears at around 200 K, which is approximately in accordance with the magnetic transition observed in LTEM. This reminds us that the magnetic transition from helical to cycloidal spin order occurring at low temperature may be triggered by the change in the easy magnetization plane from the *ab* plane at room temperature to the *bc* plane below 230 K, which may originate from the larger changes in magneto-structural effects owing to the variation in Mn-Mn distance and bond angles[23]. Although a similar evolution of the magnetic moment has been successfully modelled in other systems[45, 46], further sophisticated characterization and theory are desirable to elucidate the details of the magnetic state transition in symmetric helimagnet MnCoSi.

## 3. Conclusion

In conclusion, the intriguing magnetic structure in the non-chiral centrosymmetric helimagnet MnCoSi was investigated by LTEM. The helical spin order of ~3.2 nm in MnCoSi at ambient temperature was validated with characterizations for specimens with different orientations and serial tilting experiments. The in situ cooling LTEM experiments together with NPD measurement disclosed a low-temperature magnetic structure, namely, a cycloidal order (~1.8 nm), which retains the propagation along the *c* axis but changes the rotation plane into the *bc* plane. The discovery of a variable nanometric spin order can provide insight into the fascinating mechanism of the centrosymmetric helimagnet MnCoSi and provide a promising material for the application of spintronic devices and emergent inductance.

**Experimental Section**

***Sample preparation.*** Polycrystalline MnCoSi was grown by the arc-melting method as described in an earlier report[42]. Single crystals of MnCoSi were grown through the Sn flux method with a molar ratio Mn:Co:Si:Sn=1:1:1:15. A mixture of pure Mn, Co and Si and Sn was placed in an aluminium oxide tube and then sealed in an evacuated quartz tube. It was heated to 1423 K for 100 h and then cooled slowly to 973 K at a rate of 2 K/h. The excess Sn was removed by spinning the tube in a centrifuge at 973 K. Shiny single crystals with a size of 40×40×570 μm were obtained.

***Structure and magnetic measurements.*** The crystal structure was characterized by powder X-ray diffraction (XRD, Rigaku), energy-dispersive X-ray spectroscopy with scanning electron microscopy (SEM, FEI Quanta 250F) and transmission electron microscopy (JEOL ARM200F). The quality of single crystals was characterized by single-crystal XRD (Brucker D8 Advance). *AC* magnetic susceptibility measurements were carried out in a superconducting quantum interference device (SQUID, MPMS XL-7). In situ variable-temperature NPD measurements ($\lambda$=1.622 Å) in the heating process were carried out on the Wombat beamline at the OPAL facility of the Australian Nuclear Science and Technology Organization (ANSTO). The magnetic structure refinements for all NPD data were analysed by Fullprof software. BASIREPS in Fullprof was utilized for the magnetic representational analysis.

***Lorentz electron transmission microscopy.*** The crystalline orientation of the single crystal was further determined from electron back scatter diffraction (EBSD, ZEISS Merlin compact). Thin

LTEM specimens along the [100], [110] and [010] zone axes were fabricated from a single crystal with the desired orientation by a focused ion beam (FIB) milling apparatus. The magnetic domain was observed with a JEOL 2100F transmission electron microscope equipped with a double tilted holder, cooling holder (133 K to 298 K) and high tilted holder (±70°). Three images with under-focus, over-focus and accurate (or zero) focus were recorded by a charge-coupled device (CCD) camera in LTEM. To accurately determine the period of spiral order, selected area electron diffraction (SAED) was performed at camera lengths of 780 cm and 470 cm. The simulation of the Lorentz-TEM images was performed by means of a homemade plugin written in Digital Micrograph script.[47]

**Supporting Information**

Supporting Information is available from the Wiley Online Library or from the author.

**Acknowledgements**

This work was supported by the National Key R&D Program of China (Grant No. 2017YFA206303), the National Natural Science Foundation of China (Grant Nos. 11874410 and 11974406), the Strategic Priority Research Program (B) of the Chinese Academy of Sciences (CAS) (XDB33000000), and the China Postdoctoral Science Foundation (No. 2021M693365). We thank Ying Zhang for assistance with LTEM experiments. Zhenxiang Cheng thanks the Australian Research Council for support (DP190100150).

## Conflict of Interests

The authors declare no competing interests.